\documentclass{article}
\usepackage[utf8]{inputenc}
\usepackage{authblk}

\usepackage[verbose=true,letterpaper]{geometry}
  \newgeometry{
    textheight=9in,
    textwidth=5.5in,
    top=1in,
    headheight=12pt,
    headsep=25pt,
    footskip=30pt
  }

\title{Deep Variation Prior: Joint  Image Denoising and Noise Variance Estimation without Clean Data}
\date{}
\author[]{Rihuan Ke}
\affil[]{School of Mathematics, University of Bristol, Bristol, UK}

\usepackage{amsmath,amsfonts}
\usepackage{algorithmic}
\usepackage{algorithm}
\usepackage{array}
\usepackage[caption=false,font=normalsize,labelfont=sf,textfont=sf]{subfig}
\usepackage{textcomp}
\usepackage{stfloats}
\usepackage{url}
\usepackage{verbatim}
\usepackage{bm}
\usepackage{enumitem}
\usepackage{graphicx}
\usepackage{cite}
\usepackage{booktabs}
\usepackage{makecell}

\usepackage{appendix}

\def\qut#1{\left(#1\right)}

\def\bmy{\bm{y}}
\def\bmyh{\bm{\hat{y}}}
\def\bmx{\bm{x}}
\def\bmz{\bm{z}}

\def\bmeh{\bm{\hat{e}}}
\def\bmp{\bm{p}}
\def\bmq{\bm{q}}

\def\bmn{\bm{n}}
\def\bmnh{\bm{\hat{n}}}

\def\bbE{\mathbb{E}}

\def\qut#1{{\left( #1 \right)}}
\def\qutc#1{{\left\{ #1 \right\}}} %
\def\qutn#1{{\left\| #1 \right\|}} %
\def\quts#1{{\left[ #1 \right]}} %
\def\qutb#1{{\left| #1 \right|}} %

\def\qutan#1{\langle#1\rangle}

\def\cov#1{{\rm Cov}\left(#1\right)}
\def\var#1{{\rm Var}\left(#1\right)}

\begin{document}

\maketitle

\begin{abstract}
With recent deep learning based approaches showing promising results in removing noise from images, the best denoising performance has been reported in a supervised learning setup that requires a large set of paired noisy images and ground truth for training.  
The strong data requirement can be mitigated by unsupervised learning techniques, however, accurate modelling of images or noise variance is still crucial for high-quality solutions. The learning problem is ill-posed for unknown noise distributions. 
This paper investigates the tasks of image denoising and noise variance estimation in a single, joint learning framework.
To address the ill-posedness of the problem, we present deep variation prior (DVP), which states that the variation of a properly learnt denoiser with respect to the change of noise satisfies some smoothness properties, as a key criterion for good denoisers. 
Building upon DVP, an unsupervised deep learning framework, that simultaneously learns a denoiser and estimates noise variances, is developed. 
Our method does not require any clean training images or an external step of noise estimation, and instead, approximates the minimum mean squared error denoisers using only a set of noisy images. 
With the two underlying tasks being considered in a single framework, we allow them to be optimised for each other.
The experimental results show a denoising quality comparable to that of supervised learning and accurate noise variance estimates.
\end{abstract}

\section{Introduction}
Image
denoising refers to the process of decoupling image signals and noise from the observed noisy data.
It arises from a range of real-world applications where image acquisition systems often suffer from noise degradation. Image denoising has been investigated in the last decades and remains one of the active research topics in image processing. To decouple the signals of images and noise, a typical high-quality denoising process requires either explicitly or implicitly modelling the characteristics of images and noise, the structures of which are often complex. 

Many classic image denoising approaches are designed based on assumptions of certain image priors, for example, about smoothness \cite{perona1990scale}, self-similarity \cite{buades2005non,dabov2007image}, and sparseness of image gradients \cite{rudin1992nonlinear} or wavelet domains \cite{coifman1995translation}. With proper priors, they are shown to be effective in recovering small image details such as edges or textures. Recently, supervised deep learning (e.g., \cite{zhang2017beyond, chen2016trainable}), which does not require explicit modelling of image priors but is built on the high capability of deep neural networks in describing image data, offers a promising alternative denoising solution. Typical deep neural network based denoisers are trained using pairs of ground truth and noisy images. In doing so, image distributions are captured and encoded in the deep neural networks, which output high-quality denoising results. Nevertheless, one major limitation of these approaches is the assumption that a sufficiently large set of clean images is available for training, and such an assumption can be too strong in practical applications. 

The strong requirement for clean training images can be circumvented by unsupervised learning techniques. These techniques learn to separate clean images and noise using noisy data, for example, from paired noisy observations of single images \cite{lehtinen2018noise2noise}, a set of unpaired noisy images \cite{soltanayev2018training, batson2019noise2self, ke2021unsupervised, pang2021recorrupted}, or a single image \cite{ulyanov2018deep}. Given only single noisy images, the learning problems come with more ambiguities given the fact that noise models are not readily available from the data itself, in comparison to supervised learning where noise models can be derived from the given paired noisy-clean images and hence are known during the learning process. Therefore, to access optimal denoising quality, noise models (e.g., Gaussian distributions with known variance) are often assumed or a step of noise variance estimation is required. 

Image denoising and noise variance (noise level) estimation are two closely connected tasks. An accurate estimate of noise variance is naturally a good criterion for justifying if the size of noise being removed is proper, whereas a wrong estimate can lead to unexpected denoising effects such as residual noise or over-smoothing the denoised images. Various noise estimation approaches have been developed for different types of noise, for example, Gaussian noise \cite{immerkaer1996fast, donoho1994ideal, pimpalkhute2021digital} and Poisson noise \cite{rakhshanfar2016estimation}. In a common application setting of denoising, while the obtained noise levels can be used to configure relevant denoising algorithms, the errors in the denoised results can not be directly fed back to the noise estimation, resulting in suboptimal denoising processes. 

In the paper, we address the ill-posed problem of learning denoisers from noisy data without ground truth images or noise models given.
We proposed \textit{an end-to-end learning framework, based on deep variation prior (DVP), that approximates the minimum mean squared error (MSE) denoising objective, and that simultaneously learns a denoiser and estimates noise variances.} Specifically, we present a noise approximation based objective function that connects the MSE to a noise variance estimation problem, where errors in the noise variances lead to interpretable artefacts in denoised outputs of the learnt denoiser. 
Through DVP, which describes the smoothness properties of the variation of a learnt denoiser with respect to the change of noise, our unsupervised framework incorporates the feedback of the denoisers (artefacts) and DVP to improve both the denoisers and noise variance estimates during the learning process

The main contributions of this work are summarised as follows. 
First, we propose a regularised learning solution for denoising without clean data. A joint model is established for learning image denoisers and estimating noise variance from a set of unpaired noisy images only. 
Second, our framework provides a theoretical approximation to the mean squared error function for denoising with single observations and under unknown noise variance. With such an approximation property, it can be extended to handling other inverse imaging problems where noise-free measurements are not available. 
Third, we introduce deep variation prior, a criterion for adjusting both noise variances and denoisers based on the learnt denoisers during the learning process, and for combining deep nets' knowledge for noise variance inference and avoiding training multiple denoising models due to uncertain noise variances. 
Finally, we report experimental results showing accurate noise variance estimation as well as denoising with a quality comparable to supervised methods.

\section{Related Works}
Our work is relevant to both model-based and data-driven denoising, as well as noise variance estimation methods. We review related works in these areas. 

\medskip
\textit{\hbox{Nonlinear denoisers.}}
Nonlinear denoisers are fundamental for image denoising. Classic model-based denoisers in this category include Total Variation models \cite{rudin1992nonlinear}, BM3D \cite{dabov2007image}, Non-Local means \cite{buades2005non} and Wavelet methods \cite{coifman1995translation}. Prior knowledge about the underlying images to be restored is a core element of these approaches. Another type of nonlinear denoisers is based on deep neural networks. In particular, convolutional network based models such as DnCNN \cite{zhang2017beyond} and TNRD \cite{chen2016trainable} have shown promising results in image denoising. These models are data-driven and typically trained using pairs of noisy images and ground truth.  Our work investigates prior knowledge about images and denoising operators for data-driven denoising, which leverages intrinsic structures of noisy data. 

\medskip
\textit{\hbox{Unsupervised learning for denoising.}} Unsupervised learning methods have been studied for the tasks of image denoising, and they mitigate the strong data requirement by supervised approaches. One of the early works on this line is the Noise2Noise \cite{lehtinen2018noise2noise} which requires pairs of two independent noisy observations of single images for training. This method approximates the mean squared error loss function with the noisy samples. Ground truth images are not needed but at the same time, the paired noisy samples explicitly imply information about the noise variance. The Noise2Self \cite{batson2019noise2self} offers an alternative way of unsupervised denoising that requires only a single noisy observation per image. It follows a blind-spot strategy which removes a subset of pixels of noisy images in the input during training. While noise levels are not needed during training, Noise2Self does not take the full information of the noisy images as input. The gap between Noise2Self and supervised learning can be made smaller with a post-processing technique proposed in \cite{krull2019probabilistic}, where the outputs of the denoisers are combined with the known noise distributions and inputs for more accurate results. 
Building on a partially linear structure that is learnable from single noisy images, an unsupervised framework is proposed in \cite{ke2021unsupervised} for learning denoisers without removing pixels from images. The framework requires an estimate of noise variances but information about the precise noise distribution is not needed. 
For Gaussian noise with a known noise level, the Recorrupted-to-Recorrupted \cite{pang2021recorrupted} and SURE \cite{soltanayev2018training} methods learn denoising from unpaired noisy observations of images. The proposed method in this work uses a set of single noisy observations of images without a known noise distribution, and it contains explicit modelling of noise variance in the learning process. Particularly, with the deep variation priors, we estimate simultaneously the denoisers and the noise variance from the noisy data. The proposed model is end-to-end and network-architecture agnostic.  

\medskip
\textit{\hbox{Noise variance estimation.}}
Understanding the noise variance in noisy images is a key step for various image processing tasks. Many existing noise variance estimation (NVE) approaches are developed by assuming Gaussian white noise where the noise is signal-independent. Similar to denoising, NVE needs prior information about the images and the noise. Based on the different behaviours of noise and signal in the wavelet domain, noise variance can be estimated by selecting proper wavelet coefficients \cite{donoho1994ideal}. 
Pimpalkhute et al. \cite{pimpalkhute2021digital} propose to use edge detectors to remove edge information from the wavelet domain and improve the estimation accuracy using polynomial regression. 
NVE can also be done by using spatial filtering, for example, with the  Laplacian matrices that capture local structural information \cite{immerkaer1996fast}.
In the signal-dependent noise setting, NVE is more complex given the fact that the noise variance varies across the pixels. Foi et al. \cite{foi2008practical} develop a two-stage approach that first estimates local noise standard deviations on a set of selected regions and then fits a global parametric model for the noise variance. A patch-based method is proposed in \cite{rakhshanfar2016estimation} where patches are selected and clustered based on intensity and variances. Regression analysis is then carried out based on the noise variances computed within each cluster. 
NVE in this work, in contrast, is built upon the feedback of denoisers, and noise variances are iteratively updated according to the denoising quality.

\medskip
\textit{\hbox{Regularisation parameter selection.}} Regularisation parameter selection (RPS) has been used for finding denoising solutions with balanced data-fitting and regularisation in variational approaches. One type of method for RPS is based on Morozov’s discrepancy principle \cite{bonesky2008morozov} which seeks regularisation parameters such that the distance between the obtained solution and the noisy image matches the underlying noise variance. The L-curve criterion \cite{hansen1993use} provides an alternative approach for RPS, which chooses the optimal parameters near the corner of the L-curve in the plot of the regularisation term verse the data-fitting term. 
With a similar strategy to these approaches, the parameters of the proposed joint models are decided based on the denoised results such that they meet the relevant denoising criteria.

\section{Problem Formulation and the Main Approach} 
In image denoising, one aims to recover a clean image $\bmx$ given its noisy version $\bmy \in \mathbb{R}^m$, from the noise corruption model
\begin{equation}\label{eq:observatoinmodel}
    \bmy = \bmx + \bmn
\end{equation}
where $m$ is the number of pixels, and $\bmn$ represents the noise vector. Both $\bmx$ and $\bmn$ are random vectors following some underlying distributions. 
In this work, we assume $\bmn$ has \textit{zero mean}, and it can be either dependent or independent of $\bmx$. The distribution and variance of $\bmn$, however, are unknown. 

\subsection{The learning problems}

To find a solution for \eqref{eq:observatoinmodel}, we aim to find a denoiser $R: \mathbb{R}^m \rightarrow \mathbb{R}^m$ that takes $\bmy$ to $\bmx$. The denoiser $R$ can be learnt from data, if a set of samples are available. 

\medskip
\textit{Paired ground truth and noisy images.} The most basic scenario is supervised learning, where we assume that a set of noisy samples $y^{(i)}$ and their associated clean images $x^{(i)}$ are given. Mathematically, the denoising learning can be formulated as the problem of obtaining $R$ from the data:
\[
\{(y^{(i)},x^{(i)})\} \rightarrow R, \tag{\textbf{P1}}
\]
where we expect $R(y^{(i))}$ to be equal or similar to $x^{(i)}$. 

In general, the exact recovery of $\bmx$ from $\bmy$ is often infeasible due to the undetermined problem of \eqref{eq:observatoinmodel}, and an ideal learning process is to find an estimate $R\qut{\bmy}$ whose distance to $\bmx$ is minimal. A typical measure of their distance is the $l_2$ distance, and this leads to the mean squared error (MSE) of $R$ defined by 
\begin{equation}\label{eq:mse}
    \mathcal{L}(R) = \bbE\qut{\qutn{R\qut{\bmy} - \bmx}^2},
\end{equation}
where the expectation is taken over $\bmx$ and $\bmn$. The minimiser of MSE gives the conditional mean estimator (also known as minimum MSE estimator) $\bbE(\bmx\!\mid\!\bmy)$. In many practical situations, however, samples of clean image $\bmx$ are not given.

\medskip
\textit{Paired noisy images.} 
Instead of relying on clean images for solving the problem (\textbf{P1}), one can estimate $R$ from paired noisy samples. Given independent observations $y^{(i,1)}$, $y^{(i,2)}$ of a single image $\bmx$, the learning problem is
\[
\{(y^{(i,1)},y^{(i,2)})\} \rightarrow R. \tag{\textbf{P2}}
\]
This problem has been investigated in the Noise2Noise approach \cite{lehtinen2018noise2noise}, which provides an estimate of the conditional mean $\bbE(\bmx\!\mid\!\bmy)$. While ground truth images are not necessary here, the problem (\textbf{P2}) can be addressed with the information of noise distributions encoded in the paired observations. 
As an example, $(y^{(i,1)}-y^{(i,2)})^2/2$ provides an unbiased estimate of the noise variance.

\medskip
\textit{Single noisy images.} Further relaxation of the condition in the problem (\textbf{P1}) is to use solely a set of single noisy images. The learning problem, in this case, is described by
\[
\{y^{(i)}\} \rightarrow R, \tag{\textbf{P3}}
\]
where elements in the set $\{y^{(i)}\}$ are not linked explicitly to each other. 
Given only $\{y^{(i)}\}$, the learning problem (\textbf{P3}) is ill-posed as there exist infinitely many ways of interpreting the data. 

In this work, we focus on the problem (\textbf{P3}). The aim is to develop a regularised learning solution with denoising quality similar to that of the supervised approaches.

\subsection{Noise approximation learning objectives}

Given samples of $\bmy$, if there exists a denoiser $R$ such that the reconstructed image looks similar to a clean one, i.e., $R\qut{\bmy} \approx \bmx$, then it implies an approximation to the noise
\[
R_n\qut{\bmy} := \bmy - R\qut{\bmy} \approx \bmn.
\]
Formally, the problem of minimising the MSE \eqref{eq:mse} is equivalent to minimising
\begin{equation}\label{eq:mse-noise}
    \mathcal{L}(R) = \bbE\qut{\qutn{R_n \qut{\bmy} - \bmn}^2},
\end{equation}
where we expect the output of $R_n\qut{\cdot}$ to follow a similar distribution to the noise. 

The objective function \eqref{eq:mse-noise} reformulates the denoising problem as a noise estimation problem. From the viewpoint of \eqref{eq:mse-noise}, the correct amount of noise to be removed is characterised by the noise vector $\bmn$, which reflects the true noise distribution.

One of the main obstacles to applying \eqref{eq:mse-noise} in denoiser learning is that, in the unsupervised setting of (\textbf{P3}), neither the noise vector nor the noise distribution is accessible. This is in contrast to supervised learning where such information can be directly inferred from the noisy-clean pairs. To handle the unknown elements in the noise distribution, we start from a variant of \eqref{eq:mse-noise} by relaxing the requirement of an explicit $\bmn$, and then based on this, develop a joint learning framework for both denoising and noise variance estimation.

\medskip
\textit{\hbox{Noise approximation without $\bmn$.}} To derive a variant of \eqref{eq:mse-noise}, let $\bmz$ be a random vector satisfying 
\begin{equation}\label{eq:z}
    \bbE\qut{\bmz \mid \bmn, \bmx} = 0, \quad \cov{\bmz \mid \bmx} = M \cov{\bmn \mid \bmx},
\end{equation}
where $\cov{\bmz \mid \bmx} := \cov{\bmz,\bmz \mid \bmx }$ denotes the covariance matrix of $\bmz$ conditioned on $\bmx$, and ${M}$ is a matrix independent of $\bmn$ and conditional on $\bmx$. The definition of $\bmz$ is general because of the following reasons. 
\begin{itemize}[topsep=0pt,noitemsep]
    \item The vector $\bmz$ is not restricted to a specific distribution (and is possibly dependent on $\bmx$ and $\bmn$).
    \item $\cov{\bmz \mid \bmx}$ can be an arbitrary covariance matrix, and $M$ always exists as long as $\cov{\bmn \mid \bmx}$ is non-singular. 
\end{itemize}
Associated with the auxiliary random vector $\bmz$, a new loss function of $R$ is defined without $\bmn$ by
\begin{equation}\label{eq:Lhat}
    \widehat{\mathcal{L}}\qut{R} = \bbE\qut{ \qutn{R\qut{\bmy + \alpha \bmz} - (\bmy - \bmz/\alpha)}^2  },
\end{equation}
where $\alpha$ is a small positive number, and the expectation is taken over $\bmy$ and $\bmz$. The definition of $\widehat{\mathcal{L}}$ does not require an explicit expression of $\bmx$ or $\bmn$, but we will show next that it provides an approximation to \eqref{eq:mse-noise} if $R$ has a {partially linear structure} \cite{ke2021unsupervised}. 

\textbf{Definition 1 (partially linear structure). } A denoiser $R: \mathbb{R}^m \rightarrow \mathbb{R}^m$ has a partially linear structure if it can be decomposed into 
\begin{equation}\label{eq:pld}
    R\qut{\bmyh} = g(\bmx) + L {\bmnh} + \bmeh,
\end{equation}
where $\bmnh = \bmn + \alpha \bmz$, $\bmyh = \bmy + \alpha \bmz$, $\alpha$ is a constant, $g$ is a linear or nonlinear function, $L$ is a linear operator independent of $\bmnh$, and $\bmeh$ has a small variance. If $R$ has a partially linear structure, then we call it a partially linear denoiser. 

The set of partially linear denoisers has been shown to contain good denoisers \cite{ke2021unsupervised}. In particular, given small $\bmeh$ and bounded operator $L$, \eqref{eq:pld} defines an operator stable with respect to the change of noise. Additionally, the set of partially linear denoisers allows us to approximate the noise without knowing the exact distribution of noise, as stated by the following theorem. 

\textbf{Theorem 1 (noise approximation). } Let $R$ be a denoiser satisfying \eqref{eq:pld} with $\bmeh = 0$ and assume that $\bmn$ has zero mean, then $\widehat{\mathcal{L}}$  (defined in \eqref{eq:Lhat}) satisfies
\begin{equation}\label{eq:th1}
\widehat{\mathcal{L}}\qut{R} = \bbE\qut{ \qutn{R_n\qut{\bmyh}-M\bmn}^2 } + c,
\end{equation}
where $R_n\qut{\bmyh}:= \bmy - R\qut{\bmyh}$, $c$ is a constant independent of $R$, and the expectation is taken over $\bmz$, $\bmx$ and $\bmn$.

\textit{Proof.} By the definitions of $\widehat{\mathcal{L}}$ and  $R_n\qut{\bmyh}$, 
\begin{equation}\label{eq:mse-split}
\begin{split}
 \widehat{\mathcal{L}}(R)
  = \bbE\qut{\qutn{R_n\qut{\bmyh} - \bmz/\alpha}^2} 
=  \bbE\qut{\qutn{\qut{R_n\qut{\bmyh} - M \bmn} + \qut{M \bmn - \bmz/\alpha} }^2 }. 
\end{split}
\end{equation}
According to \eqref{eq:pld}, the first term on the right hand side of \eqref{eq:mse-split} can be rewritten as 
\[
R_n\qut{\bmyh} - M \bmn = \bmx - g(\bmx) - L \bmnh + (\bmn-M\bmn). 
\]
Denoting by $\qutan{\cdot,\cdot}$ the inner product operator, we will show that the cross term \\ $\bbE\qut{\qutan{R_n\qut{\bmyh} - M \bmn, M \bmn - \bmz/\alpha}}$ of \eqref{eq:mse-split} is constant in the following three steps. 

(i). By the assumption $\bbE\qut{\bmn \mid \bmx} = 0$ and $\bbE\qut{\bmz \mid \bmx} = 0$, 
\[
\bbE\qut{\qutan{\bmx - g(\bmx), M \bmn - \bmz/\alpha}} = 0.
\] 

(ii). It follows from the assumption $\bbE\qut{\bmz \mid \bmn} = 0$ that 
\[
\begin{split}
    \bbE\qut{\qutan{L\bmnh, M \bmn - \bmz/\alpha}}  
    =  & \bbE\qut{\qutan{L\bmn + \alpha L \bmz, M \bmn - \bmz/\alpha}} \\
    = & \bbE_{\bmx} \qut{ {\rm tr}\qut{\cov{L\bmn, M\bmn} \mid \bmx} - {\rm tr}\qut{\cov{L\bmz, \bmz \mid \bmx}} } \\
    = & \bbE_{\bmx} \qut{ {\rm tr}\qut{L \cov{\bmn \mid \bmx } M^T } - {\rm tr}\qut{L \cov{\bmz \mid \bmx}} } 
    =  0, \\
\end{split}
\]
where $M^T$ denotes the transpose of $M$, ${\rm tr(\cdot)}$ is the trace of a matrix, and the second and the last equalities follow from the condition \eqref{eq:z}. 

(iii). Let $c_1:=\bbE\qut{\qutan{\bmn-M\bmn, M \bmn - \bmz/\alpha}}$. This together with (i) and (ii) gives
\begin{equation}\label{eq:4221}
\begin{split} \bbE\qut{\qutan{R_n\qut{\bmyh} - M \bmn, M \bmn - \bmz/\alpha}} 
=   \bbE\qut{\qutan{\bmn-M\bmn, M \bmn - \bmz/\alpha}} = c_1,    
\end{split}    
\end{equation}
which is independent of $R$. 

Finally, let $c_2 := \bbE\qut{\qutn{{M \bmn - \bmz/\alpha}}^2}$, and $c:=c_1 + c_2$, then
\eqref{eq:mse-split} and \eqref{eq:4221} give \eqref{eq:th1}. \hfill{$\square$}

\textbf{Remark 1.} Comparing \eqref{eq:th1} with \eqref{eq:mse-noise}, we observe that \eqref{eq:th1} is a general form of \eqref{eq:mse-noise}, whereas \eqref{eq:th1} is connected to \eqref{eq:Lhat} which is free of $\bmn$-related terms. 
By Theorem 1, $\widehat{\mathcal{L}} - c$ is an approximation to $\mathcal{L}$ given in \eqref{eq:mse-noise} when $M$ is close to the identity matrix $I$. 

\textbf{Remark 2.}
It can be seen that \eqref{eq:th1} holds without regard to the noise types, and $\widehat{\mathcal{L}}$ depends on the covariance of $\bmz$ but not its distribution. 

\textbf{Remark 3.} If $\bmeh \neq 0$, then \eqref{eq:th1} does not hold. However, if $\bbE\qut{\qutn{\bmeh}^2}$ is small, as required by the partially linear structure, then the distance between $\widehat{\mathcal{L}}\qut{R}$ and $\bbE\qut{ \qutn{R_n\qut{\bmyh}-M\bmn}^2 } + c$ is small. 

By Theorem 1, to obtain a denoiser that removes the correct amount of noise, we aim to have $M=I$ (i.e., $\bmn$ and $\bmz$ have the same covariance), and under the condition \eqref{eq:pld}, minimising $\widehat{\mathcal{L}}$ is equivalent to finding $R_n\qut{\bmyh}$ that best matches the noise $\bmn$ in the mean squared distance. For $M \neq I$, $R_n\qut{\bmyh}$ is fitted to a variant of $\bmn$ and consequently, $R$ may not output clean images. The matrix $M$, however, is unknown since $\cov{\bmn \mid \bmx}$ is not given.

\textit{In this work, the formulations \eqref{eq:Lhat} and \eqref{eq:th1} provide a theoretical foundation for  unsupervised denoising and noise variance estimation.} Importantly, the objective functions are less data-dependent as they are defined without clean images or being tied to specific noisy types, and they imply a connection between the noise (possibly with an inaccurately estimated variance which implies $M \neq I$) and the computed denoisers. The unknown noise variance is reflected in the denoised outputs, and so are the unknown noise components. 
With the deep variation priors introduced in the next subsection, we circumvent the unknown noise variances and exploit the above connection for unsupervised denoising. 

\subsection{Deep variation prior}\label{subsec:dvp}
The matrix $M$ is connected to $R_n$, based on the fact that the minimisation of the objective function \eqref{eq:th1} motivates a noise estimation satisfying 
\begin{equation}\label{eq:4151}
    R_n\qut{\bmyh} \approx M \bmn,
\end{equation}
where $R_n$ (or the denoiser $R$) can be represented by a parameterised function, typically a deep neural network.  
Recall that matrix $M$ on the right hand side is unknown. 
However, for small $\alpha$ (hence $\bmyh \approx \bmy$), \eqref{eq:4151} implies an approximate linear correspondence (defined by $M$) between the outputs of $R_n$ and the noise, conditioned on $\bmx$. To have a closer look at the correspondence, we define the variation of denoisers.

\textbf{Definition 2 (variation of denoisers). } Let $R$ be a denoiser as an operator from $\mathbb{R}^m$ to $\mathbb{R}^m$. The variation of $R$ with respect to noise is defined as
\[
\delta R := R(y^{(2)}) - R(y^{(1)})
\]
where $y^{(1)}$ and $y^{(2)}$ are noisy versions of a single image. 

Similarly, the variation of $R_n$ is defined as $\delta R_n = R_n\qut{y^{(2)}} - R_n\qut{y^{(1)}} = (y^{(2)} - y^{(1)}) + (R(y^{(1)}) - R(y^{(2)})) = \delta y - \delta R$, where $\delta y := y^{(2)} - y^{(1)}$. Here for the simplicity of illustration, we take the auxiliary vector $\bmz$ as zero. 
With the definition of $\delta R$, for each image $\bmx$, the condition \eqref{eq:4151} implies 
\begin{equation}\label{eq:871}
\delta R \approx \delta y - M \delta y.
\end{equation}
The variation of $R$ encodes information of $M$. For instance, if $M = \rho I$ for $\rho < 1$, i.e., the noise variance is underestimated, then $\delta R \approx (1-\rho) \delta y$ contains some residual components of $\delta y$. 

In comparison with \eqref{eq:4151}, the formulation \eqref{eq:871} does not involve explicit expressions of $\bmn$ (which is unknown) and indicates an approximation of the matrix $M$ (and hence $\cov{\bmn \mid \bmx}$) from $\delta R$ and $\delta y$. 
Notice that the two sides of the approximate equality are not equal in general, because the noise usually can not be completely separated from noisy images (i.e., $\bbE( \qutn{R_n\qut{\bmyh}-M\bmn}^2) \neq 0$). 
However, the components of $\delta y - M \delta y$ still play a role in $\delta R$, subject to the uncertainty in the noise and image separation. The connection between $\delta R$ and $M$ can be better characterised by taking into account the uncertainty.

\begin{figure*}[ht]
    \centering
    \setlength{\tabcolsep}{10pt}
    \makebox[\textwidth][c]{
    \begin{tabular}{ccc}
        \begin{tabular}{c}
            \includegraphics[width=0.20\linewidth]{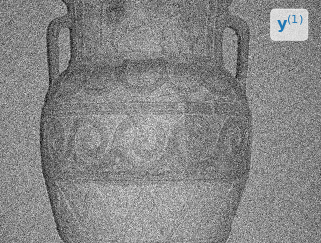} \\
            \includegraphics[width=0.20\linewidth]{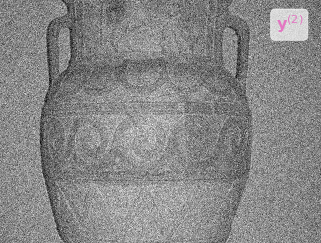}
        \end{tabular}
        & 
        \begin{tabular}{c}
            \includegraphics[width=0.20\linewidth]{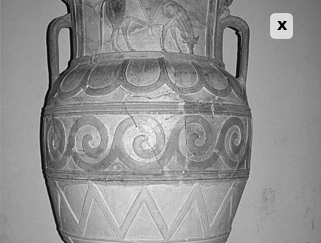} \\
            \includegraphics[width=0.20\linewidth]{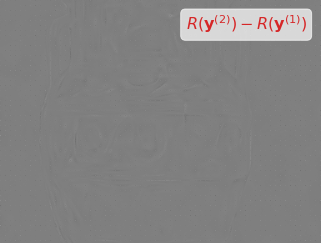} 
        \end{tabular}
        & 
        \begin{tabular}{c}
            \includegraphics[trim=10 20 12 0, clip, width=0.35\linewidth]{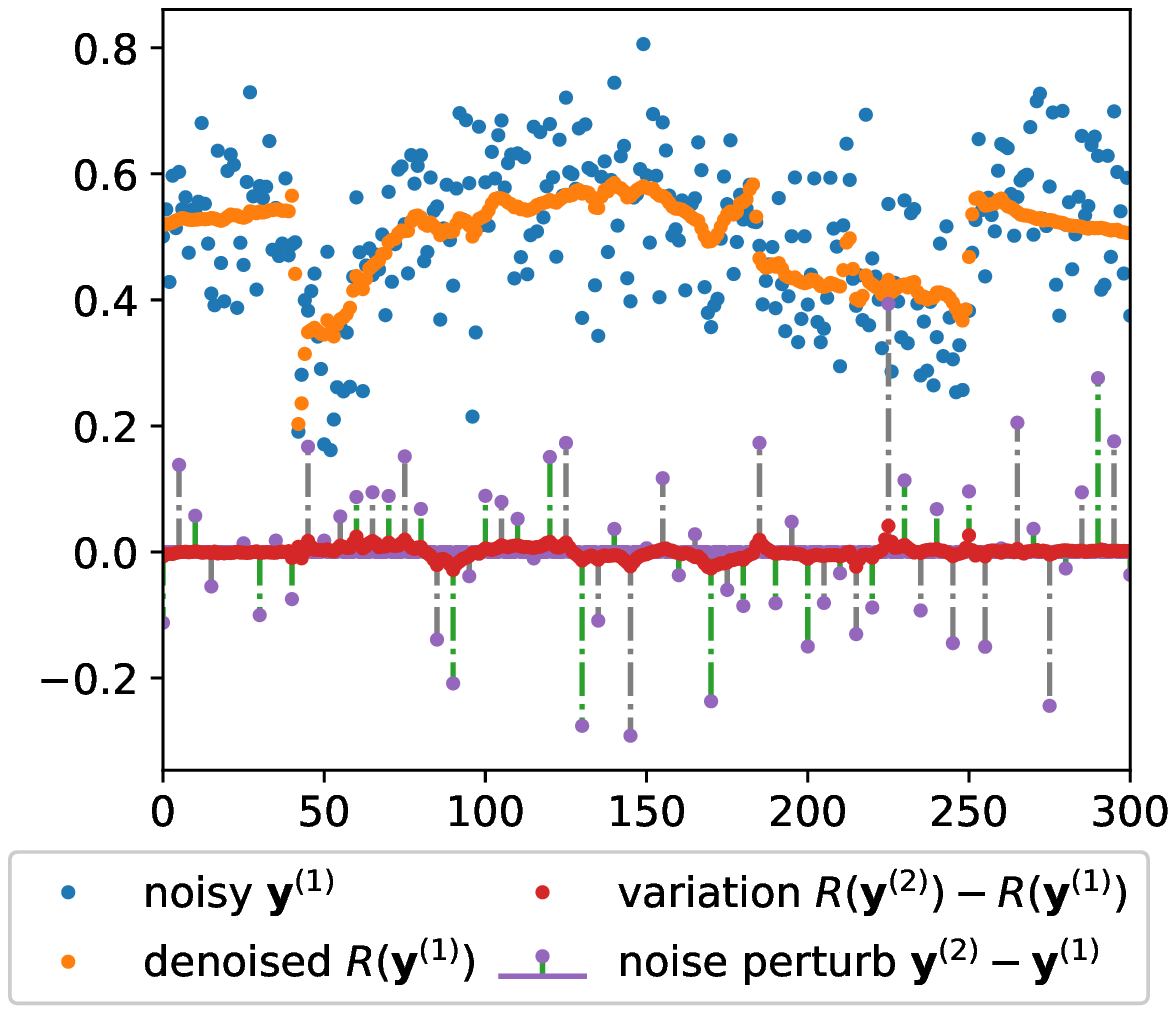}
        \end{tabular}
        \\
        (a). Noisy images & (b). GT and variation & (c).  reconstruction (standard noise variance) \\
    \end{tabular}}
    \\
    \bigskip
    \makebox[\textwidth][c]{
    \begin{tabular}{cc}
        
         \begin{tabular}{c}
            \includegraphics[trim=10 20 12 0, clip, width=0.35\linewidth]{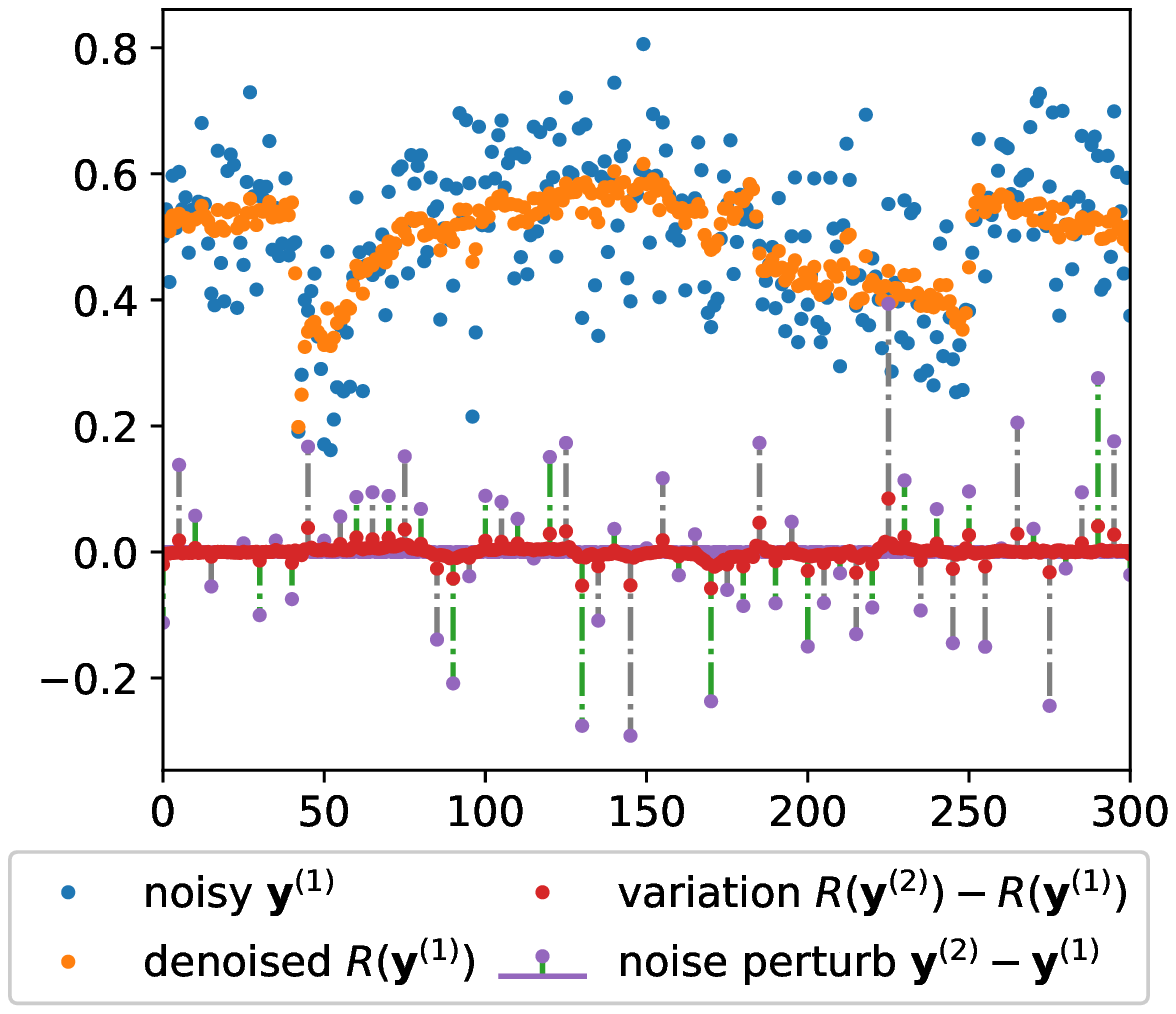} 
        \end{tabular}
        &
         \begin{tabular}{c}
            \includegraphics[trim=10 20 12 0, clip, width=0.35\linewidth]{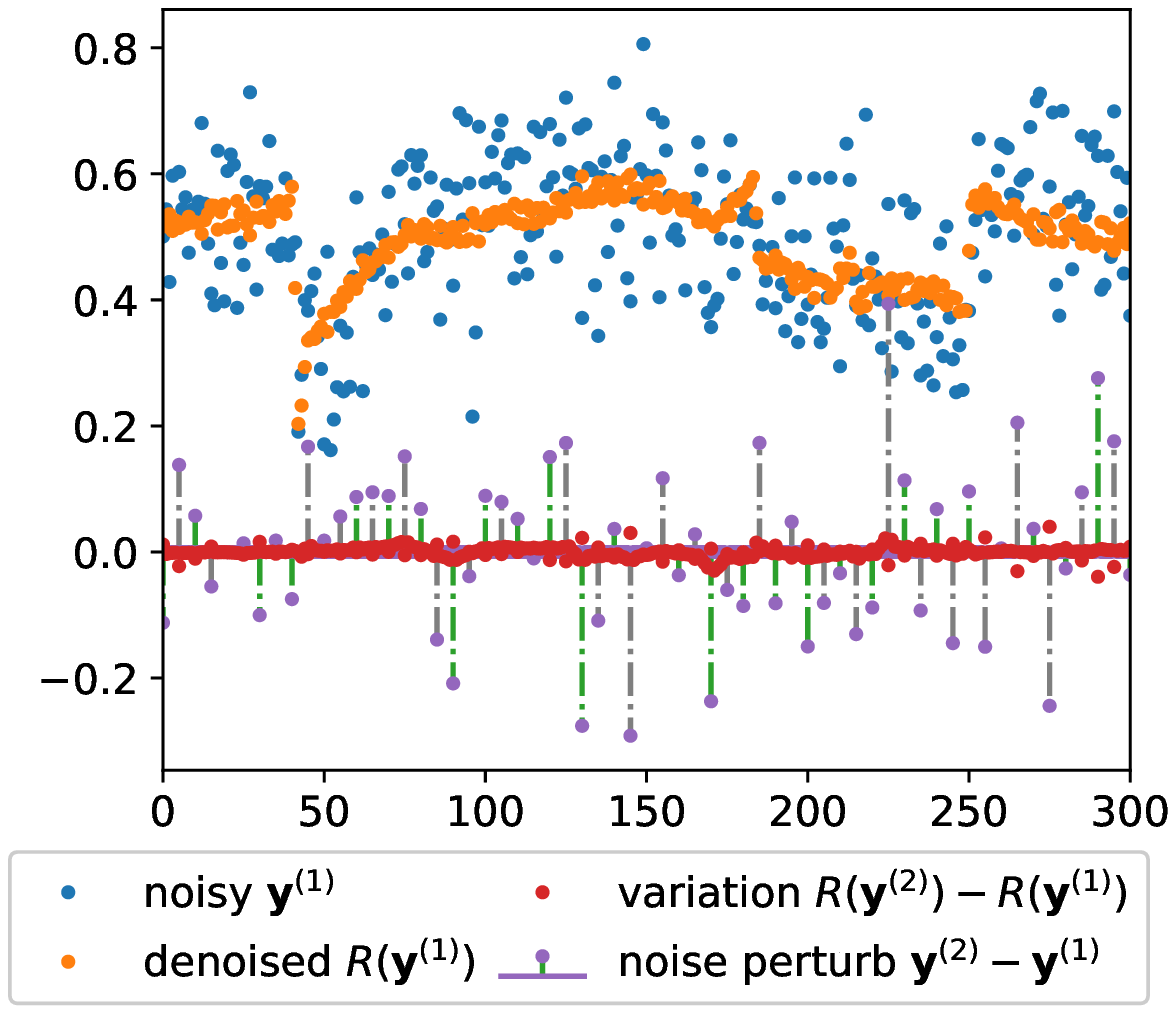}
        \end{tabular}
        \\
        (d). reconstruction (underestimated noise variance) &
        (e). reconstruction (overestimated noise variance) \\
    \end{tabular}} \\

    \caption{\textit{Deep variation prior reflects how the outputs of a denoiser change as the noise changes}. Best view in colour. Given a noisy image $\mathbf{y}^{(1)}$ and its perturbed version $\mathbf{y}^{(2)}$ (with different noise) in (a), the \textit{variation of a denoiser $R$} is an image defined as $\delta R:= R(\mathbf{y}^{(2)})-R(\mathbf{y}^{(1)})$ (as shown at the bottom of (b)). The red dots in the plots in (c)-(e) illustrate how the variation  $\delta R$ is connected to the noise change $\delta y:=\mathbf{y}^{(2)} - \mathbf{y}^{(1)}$ on one row of the image: (c) is the result of a denoiser that is trained with the true noise variance, and the variation (represented by the red dots) is smooth except in a small set of locations. The plots in (d) (resp. (e)) show the output of a denoiser that is learnt with underestimated (resp. overestimated) noise variance, and in contrast to (c), it is less smooth and has residual noise. 
    }
    \label{fig:demo}
\end{figure*}

We introduce \textit{deep variation prior} (DVP), as a prior for $\delta R$ and more generally a criterion for denoising. The main idea is that, \textit{for a learnt denoiser $R$ that correctly approximates the clean images, the variation $\delta R$ is a piece-wise smooth function.} Such a prior can be interpreted from the following two perspectives. 
\begin{itemize}[topsep=0pt, noitemsep]
    \item For denoiser $R$ that correctly predicts smooth regions of the image, the predictions are expected to be minimally dependent on the changes of noise, and hence the variation $\delta R$ should be smooth on these regions as well. 
    \item 
    Both images and noise have non-smooth components. However, if $M$ is equal to the identity matrix $I$, then both image structures and noise structures are removed from the right-hand side of \eqref{eq:871}. 
    Therefore, the minimisation of $\widehat{\mathcal{L}}$ in \eqref{eq:th1} tends to remove these structures from $\delta R$. 
\end{itemize}
Particularly, the variation $\delta R$ does not necessarily equal zero for $M=I$, but it is described as a function with smoothness properties, in contrast to the right hand side of \eqref{eq:4151} which is identically zero. 

The concept of DVP for learning denoisers in the unsupervised setting with estimated noise variances is illustrated in Figure \ref{fig:demo}, which contains examples of $\delta R$. The denoiser learnt with the true noise variance leads to a variation image that is smooth except on a small set of locations. The variations of denoisers that are learnt with overestimated (or underestimated) noise variance are not smooth. Intuitively, if $M \neq I$, then parts of the noise will remain in $\delta R$ according to the relationship \eqref{eq:871}. Typical non-smooth components of $\delta y$ lead to non-smooth noise patterns in $\delta R$ which violates the criterion of DVP. Therefore, to obtain a more accurate description of the relationship in \eqref{eq:871}, the variation $\delta R$ can be decomposed into two parts:
\begin{equation}\label{eq:deltaR-decompose}
  \delta R = \delta  R^\parallel + \delta R^\perp
\end{equation}
where $R^\parallel$ is the learnt denoiser associated with the ideal case of $M=I$, and $R^\perp$ corresponds to the residual components of noise $(I-M)\delta y$ in $R$ (violating DVP), from which we estimate $M$ and learn better denoisers. 
In summary, in the unsupervised denoising approach described in the next subsection, the aim is to learn a denoiser by minimising $\widehat{\mathcal{L}}$ given in \eqref{eq:Lhat} and removing $\delta R^\perp$ (or equivalently, letting $M=I$).

\subsection{Summary of the proposed method}\label{subsect:summaryproposedmethod}
This subsection contains a summary of the proposed method, followed by the details and specifications of the learning framework presented in the next section. 

Given a set of samples of noisy images $\bmy$ and in the absence of both the ground truth images and known noise distributions, our method finds {a denoiser} through \eqref{eq:Lhat} (an approximation to the MSE) and {an estimate of the noise variance} through the random vector $\bmz$ in \eqref{eq:z}. It is an iterative method, each iteration of which consists of two main steps:

(S1) Minimisation of the loss function $\widehat{\mathcal{L}}$ over the sets of partially linear denoisers $R$, i.e., with the partial linearity constraint \eqref{eq:pld} such that $\bmeh$ is small.  

(S2) Based on the obtained $R$ and DVP, estimating the underlying matrix $M$ and using it to update the covariance of $\bmz$ such that the new $M$ is as close to the identity matrix $I$ as possible. 

The two steps (S1) and (S2) are carried out alternatingly in a single learning process. Such a process aims to find $M \approx I$, from which an estimate of noise variance can be obtained according to \eqref{eq:z}. 

Instead of raw-data based noise variance estimating as in many classic approaches, our noise variance estimation is based on the learnt denoisers, hence incorporating the clean image distributions encoded therein as well as feedback from the denoisers.
The interplay between the two components allows them to be optimised for each other.  
In other words, the updates of $R$ in (S1) enable better estimation of the noise variance in (S2), while the improved noise variance ($M$ is closer to $I$) contributes to more accurate denoisers implied by Theorem 1.

\section{Deep learning framework for denoising and noise variance estimation}

In the rest of the paper, we focus on pixel-wise independent noise, which simplifies the structure of $M$. Unless specified otherwise, we assume that
\begin{itemize}[leftmargin=1cm]
    \item[\textbf{(A1)}] The entries of the noise $\bmn$ are zero-mean and independent random variables, and $\bmn_i$ (the $i^{\rm th}$ entry of $\bmn$) is conditional on $\bmx_i$ but \textit{independent} of pixel locations.
    \item[\textbf{(A2)}] The vector $\bmz$ is zero-mean and pixel-wise independent conditioned on $\bmx$ and $\bmn$.
\end{itemize}
Therefore, $\cov{\bmz \mid \bmx}$, $\cov{\bmn \mid \bmx}$ and $M$ are diagonal matrices. Let the conditional variance of $\bmz_i$ be denoted by $\var{\bmz_i \mid \bmx_i}$, then
\[
\cov{\bmz \mid \bmx} = {\rm diag}({\var{\bmz \mid \bmx}})
\]
where ${\var{\bmz \mid \bmx}}$ represents the vector of $\var{\bmz_1 \mid \bmx_1}$, $\cdots$, $\var{\bmz_m \mid \bmx_m}$, and ${\rm diag}(\cdot)$ converts a vector to a diagonal matrix. 
Similarly, 
$
\cov{\bmn \mid \bmx} = {\rm diag}(\var{\bmn \mid \bmx}).
$
Then the second equality in \eqref{eq:z} is equivalent to  
\begin{itemize}[leftmargin=1cm]
    \item[\textbf{(A3)}] $\var{\bmz \mid \bmx} = M \var{\bmn \mid \bmx}$. 
\end{itemize}
The diagonal entries of $M$ are $M_{ii}=M_{ii}(\bmx_i)$. For ease of presentation, we assume $\var{\bmn_i \mid \bmx_i} \neq 0$ for all $i$, hence $M$ exists for any distributions of $\bmz$. 

Next, we first formulate the variance of $\bmz$. Then we present how to minimise \eqref{eq:Lhat} with constraints based on the generated vectors $\bmz$. Finally, the methods of how to update the variance of $\bmz$ based on DVP, which includes an estimation of $M$, will be discussed. 

\subsection{Formulating $\bmz$}
Let $\bmz$ be zero-mean and pixel-wise independent (conditional on $\bmy$) with mean and variance given respectively by
\begin{equation}\label{eq:8174}
\bbE\qut{\bmz_i \mid \bmy_i} = 0, \quad {\rm and} \quad \var{\bmz_i \mid \bmy_i} = f\qut{\bmy_i},  
\end{equation}
for any pixel $i$, where $f$ is a function to be determined. %

\textbf{Remark 4.} 
By \eqref{eq:8174}, the variance of $\bmz_i$ is a function of the observed value $\bmy_i$ instead of $\bmx_i$ and $\bmn_i$, hence samples of $\bmz_i$ can be generated without knowing the clean images. 
Clearly, the above formulation of $\bmz$ satisfies \textbf{(A2)}, and \textbf{(A3)} holds for some existing $M$.
A typical choice of $\bmz_i$ is the Gaussian random variable $\mathcal{N}(0, f(\bmy_i))$. 

Our goal is therefore to find a function $f\qut{\cdot}$ such that $M$ is as close to $I$ as possible, as required by (S2) (Cf. Subsection \ref{subsect:summaryproposedmethod}). The necessary and sufficient condition for $M=I$ is given below. 

\textbf{Proposition 2.} Assume that $\bmz$ satisfies \eqref{eq:8174} and $M$ satisfies \textbf{(A3)}. Then $M=I$ if and only if $f$ satisfies 
\begin{equation}\label{eq:conditoinM=I}
\bbE_{\bmy_i}\qut{f(\bmy_i) \mid \bmx_i } = \var{\bmn_i \mid \bmx_i}
\end{equation}
for any pixel $i$.

\textit{Proof.} By \eqref{eq:8174}, the conditional variance of $\bmz_i$ is computed as
\[
\begin{split}
\var{\bmz_i \mid \bmx_i } & = \bbE_{\bmy_i}\qut{ \var{\bmz_i \mid \bmy_i} \mid \bmx_i} \\
& = \bbE_{\bmy_i}\qut{f(\bmy_i) \mid \bmx_i},
\end{split}
\]
in which $\bbE_{\bmy_i}$ means the expectation taken over $\bmy_i$. 
This together with \textbf{(A3)} leads to a formulation of the diagonal entries of $M$:
\[
\bbE_{\bmy_i} \qut{f(\bmy_i) \mid \bmx_i } = M_{ii} \var{\bmn_i \mid \bmx_i }, \quad i=1,\cdots,m.
\]
Therefore the desired result follows from the fact that $M$ is a diagonal matrix.
\hfill{$\square$}

It is worth noting that the function $f$ satisfying  \eqref{eq:conditoinM=I} may take different forms depending on the noise distributions. For example, if $\bmn$ is i.i.d. noise, then $f$ is a constant function equal to the noise variance. For Poisson noise with variance $\var{\bmn_i \mid \bmx_i} = \lambda \bmx_i$ (where $\lambda$ is a Poisson distribution parameter), if we let $f(y) = \lambda y$, then $\bbE_{\bmy_i}\qut{f(\bmy_i) \mid \bmx_i } = \lambda \bmx_i = \var{\bmn_i \mid \bmx_i}$. 

In this work, we approximate $f$ with a low-dimensional space 
\begin{equation}\label{eq:betak}
    \left\{ f = \sum_{k=1}^K \beta_k f_k \mid \beta_k \in \mathbb{R} \right\},
\end{equation}
where $\{f_k\}$ are basis functions. The number $K$ is typically small for computation considerations. The parameters $\{\beta_k\}$ are iteratively updated, as a step of (S2). 

\subsection{Minimising $\widehat{\mathcal{L}}$}
Once $\bmz$ is defined, 
 the objective function $\widehat{\mathcal{L}}$ given in \eqref{eq:Lhat} can be evaluated. In practice, we minimise its corresponding empirical risk over the given samples of $\bmy$, which can be implemented by standard stochastic gradient descent methods. 

The partially linear structure \eqref{eq:pld} can be imposed following the penalty method proposed in \cite{ke2021unsupervised}. Specifically, if $\hat{y}^{(1)}$, $\hat{y}^{(2)}$, $\cdots$ $\hat{y}^{(P)}$ are variants of a noisy sample (associated with a single clean image) and they satisfy  $\sum_{j=1}^P \tau_j \hat{y}^{(j)} = 0$, then with the decomposition \eqref{eq:pld},
\[
\sum_{j=1}^P \tau_j R\qut{ \hat{y}^{(j)} } = \tau_j \hat{e}^{(j)}
\]
where $\hat{e}^{(j)}$ is the residual term associated with $\hat{y}^{(j)}$. A typical value of $P$ is $3$. The above equality does not require explicit formulations of the function $g$ or the linear operator $L$, and hence an explicit decomposition of $R$ is not needed. Based on these observations, one can define a penalty term $\mathcal{L}_c := \bbE_{\{\hat{y}^{(1)} \cdots \hat{y}^{(P)}\}}  \qutn{W\sum_{j=1}^P \tau_j R\qut{ \hat{y}^{(j)} }}^2$ for $\bmeh$, where $W$ is a diagonal weighting matrix for balancing the loss over the pixels. To conclude, the overall loss function is 
\begin{equation}\label{eq:L-all}
    \mathcal{L}_{\rm all} = \widehat{\mathcal{L}} + \gamma \mathcal{L}_c,
\end{equation}
where $\gamma$ is a weighting parameter. 

\subsection{Updating the variance of  $\bmz$}\label{subsect:update-z}
The variance of $\bmz$ is represented by the parameterised function $f$ in \eqref{eq:betak}. 
Let the parameters at iteration $l$ be denoted by $\beta_k^{(l)}$, then the variance of $\bmz$ is given by
\begin{equation}\label{eq:8714}
    f^{(l)}(\bmy) = \sum_{k=1}^K \beta_k^{(l)} f_k(\bmy).
\end{equation}
Here $f_k(\bmy)$ is a vector with entries $f_k(\bmy_{1}), \cdots, f_k(\bmy_{m})$.
To update $\bmz$, we compute parameters $\{\beta_k^{(l+1)}\}$.

Recall that in (S2), $\bmz$ is updated in order to meet the criterion $M \approx I$, which means $\var{\bmn_i \mid \bmx_i} \approx \bbE_{\bmy_i}\qut{f(\bmy_i) \mid \bmx_i }$ according to the condition \eqref{eq:conditoinM=I}. Motivated by this, $\{\beta_k\}$ are updated as follows. First, at iteration $l$, the variance $\var{\bmn_i \mid \bmx_i}$ can be expressed in terms of $f^{(l)}$ and $M$ (according to \textbf{(A3)} and the definition of $f^{(l)}$):
\[
\var{\bmn \mid \bmx} = M^{-1} \var{\bmz \mid \bmx} = M^{-1} \bbE_{\bmy}\qut{f^{(l)}(\bmy) \mid \bmx }.
\]
Second, combining this with \eqref{eq:conditoinM=I}, we aim to solve
\begin{equation}\label{eq:8614}
\min_{\beta_1, \cdots, \beta_K} \qutn{ \bbE_{\bmy}\qut{f(\bmy) \mid \bmx } - 
M^{-1} \bbE_{\bmy}\qut{f^{(l)}(\bmy) \mid \bmx }}
\end{equation}
to obtain new parameters $\{\beta_k^{(l+1)}\}$.
To compute $\{\beta_k^{(l+1)}\}$, one needs to estimate the matrix $M$ and the expectations, the details of which are given next. 

\medskip
\textit{\hbox{C.1) Estimating $M$ from $\delta R$.}} The matrix $M$ can be obtained from $\delta R$ in the following steps. 

First, let $R^\parallel$ be the minimizer of the MSE 
\[
\widehat{\mathcal{L}}^\parallel\qut{R} = \bbE\qut{ \qutn{R_n\qut{\bmyh}-\bmn}^2 }.
\]
The function $\widehat{\mathcal{L}}^\parallel$ is a special case of $\widehat{\mathcal{L}}$ with $M=I$. As $R^\parallel$ is obtained with the correct noise variance, we assume that $\delta R^\parallel$ satisfies the DVP (Cf. Subsection \ref{subsec:dvp}). 

Second, let $R^*$ be the minimizer of $\widehat{\mathcal{L}}$ associated with $\bmz$ whose variance is $f^{(l)}(\bmy)$ ($M$ is not necessarily equal to $I$ here). 
With $R^\parallel$ defined above, we decompose $R^*$ into the following form
\begin{equation}\label{eq:7512}
    R^* = R^\parallel + R^\perp,
\end{equation}
where $R^\perp$ contributes to the non-smooth parts of $\delta R^*$ that violate DVP. This decomposition leads to a split of $\delta R^*$ in the form of \eqref{eq:deltaR-decompose}, where $\delta R^\perp$ is a crucial component for estimating $M$, as explained in Subsection \ref{subsec:dvp} and detailed next. 

Third, 
the desired matrix $M$ is linked to $R^\perp$ by
\[
R^\perp(\bmyh) \approx g^\perp(\bmx) + L^\perp(\bmn + \alpha \bmz)
\]
where $g^\perp$ is a function, and 
\begin{equation}\label{eq:LM}
    L^\perp = (I-M) (I + \alpha^2 M)^{-1}.
\end{equation}
The details of the derivation of $L^\perp$ are given in Appendix \ref{sect:appendixA}. Notice that $M$ is a diagonal matrix, and so is  $L^\perp$. Consequently, $L^\perp$ can be estimated by $ L^\perp \delta y \approx \delta R^\perp$. 
To estimate $\delta R^\perp$, one can remove smooth components of $\delta R^\parallel$ from $\delta R$. Specific information on this is given in the experimental section \ref{sect:experiment}. 

In summary, the matrix $M$ can be computed as 
\begin{equation}\label{eq:3141}
    M = (\alpha^2 L^\perp + I)^{-1} (I-L^\perp).
\end{equation}  
Clearly, if $R^\perp = 0$, then $L^\perp = 0$ and $M = I$.

\medskip
\textit{\hbox{C.2) Estimating conditional expectations.}} 
The conditional expectations in \eqref{eq:8614} are pixel location independent, i.e., $\bbE_{\bmy_i}\qut{f(\bmy_i)\mid \bmx_i} = \bbE_{\bmy_j}\qut{f(\bmy_j)\mid \bmx_j}$ if $\bmx_i = \bmx_j$, which is a result of the assumptions \textbf{(A1)} and \textbf{(A2)}. Hence, given a sample $y$ of $\bmy$ (associated with the clean image $x$), the conditional expectations are estimated as
\begin{equation}\label{eq:9731}
\bbE_{\bmy_i}\qut{h(\bmy_i)\mid \bmx_i = x_i} \approx {\rm average} \qut{\qutc{ h(y_j) \mid x_j = x_i}} 
\end{equation}
where $h \in \{f, f^{(l)} \}$. The right hand side of the above approximation is an average over different pixel locations of a single sample. To represent this in a compact form, we define a matrix $A$, the entries of which are
\begin{align*}
    &A_{ij} = 0 \quad &{\rm if}~ {x_i \neq x_j}, \\
    &A_{ij} = 1/\qutb{\{ s \mid x_s = x_i \}} \quad &{\rm if}~ {x_i = x_j},
\end{align*}
in which $\qutb{\cdot}$ denotes the cardinality of a set. In practice, the clean image $x$ is unknown, so we approximate $A$ by replacing $x$ with its estimated version $R^*\qut{\hat{y}}$. With the definition of $A$, we can rewrite \eqref{eq:9731} as $\bbE_{\bmy}\qut{h(\bmy) \mid \bmx = x } \approx A h(y) $. 

Moreover, recall that $M$ in \textbf{(A3)} is a diagonal matrix with diagonal entries $M_{ii}=M_{ii}(\bmx_i)$. This implies that $M^{-1} = {\rm diag}\qut{AM^{-1}\mathbf{1}}$ where $\mathbf{1}$ is the vector of all ones. In the computations of \eqref{eq:8614}, $M^{-1}$ can be replaced with ${\rm diag}\qut{AM^{-1}\mathbf{1}}$ which is more robust to the errors in the computed $M$. 

Finally, at iteration $l$, the minimisation of \eqref{eq:8614} is carried out over a mini-batch of noisy images instead of the entire dataset. This, together with the approximated matrices and expectations, introduces randomness in the computed solutions. A common strategy to reduce the randomness is to let the sequence $\beta_k^{(l)}$ be an experiential moving average of the computed solutions. With this practical consideration, the new parameters of $\bmz$ are computed as:
\begin{equation}\label{eq:8615}
\begin{split}
    \{\beta_k^*\} &= \arg\min_{\{\beta_k\}} \qutn{ A \sum_k \beta_k f_k(y) - 
{\rm diag}\qut{A M^{-1} \mathbf{1}} A f^{(l)}(y)}^2 , \\
 \beta_k^{(l+1)} &= (1-\upsilon) \beta_k^{(l)} + \upsilon \beta_k^{*}, \quad k=1,2,\cdots,K, \\
\end{split}
\end{equation}
where $\upsilon>0$ is a constant. 
The algorithm converges when $R^*$ satisfies the DVP, as in this case $R^\perp = 0$ and $M=I$, and as a result of \eqref{eq:8615}, $\beta_k^{(l+1)} = \beta_k^{(l)}$. 
The overall algorithm is summarised in Algorithm \ref{alg:alg1}.

\begin{algorithm}[h]
\caption{Learning framework}\label{alg:alg1}
\begin{algorithmic}
\STATE
\STATE \textbf{Require:} parameterised denoiser $R$, $\{\beta_k^{(0)}\}$, $\gamma$. 

\STATE \textbf{Results:} learnt denoiser $R^*$, the variance of $\bmz$

\STATE

\STATE Set $l=0$ and initialise $R$.
\STATE \textbf{While} not converge \textbf{do}

\STATE \hspace{0.5cm} Selection randomly a mini-batch of noisy images $\{y\}$

\STATE \hspace{0.5cm} For each $y$, generate $\hat{y}$, $\hat{y}^{(1)}, \cdots, \hat{y}^{(P)}$. 

\STATE \hspace{0.5cm} \textbf{(S1) update $R$}

\STATE \hspace{1cm} Carry out a gradient step of $\mathcal{L}_{\rm all}$ given in \eqref{eq:L-all}

\STATE \hspace{0.5cm} \textbf{(S2) update the variance of $\bmz$}

\STATE \hspace{1cm} Compute $M$ from \eqref{eq:3141}
\STATE \hspace{1cm} Compute $\{\beta_k^{(l+1)}\}$ by minimising \eqref{eq:8615} over the mini-batch

\STATE \hspace{0.5cm} Set $l \leftarrow l+1$
\STATE \textbf{End}
\end{algorithmic}
\label{alg1}
\end{algorithm}

\section{Experiments}\label{sect:experiment}
Experiments are carried out on benchmark denoising datasets as well as real microscopy images to evaluate the performance of the proposed approach.
The rest of this section is structured as follows. First, the network architecture and training specifications, which are shared among all experiments unless specified otherwise, are given (Subsection \ref{subsect:trainingdetails}). Second, the results on benchmark denoising datasets are presented (Subsection \ref{subsect:denoising_varianceestimation}). Finally, we demonstrate the performance of the proposed approach in denoising real microscopy images where ground truth images for training are unavailable (Subsection \ref{subsect:microscopyimages}). 

\subsection{Network architectures and training specifications}\label{subsect:trainingdetails}
The proposed approach is network-architecture agnostic. The denoiser $R$ is generally defined as a parameterised operator and hence the algorithm applies to an arbitrarily selected network architecture for denoising. In all experiments, we use one of the benchmark architectures DnCNN \cite{zhang2017beyond}.   

\medskip
\textit{Optimisation.} The minimisation of the loss function \eqref{eq:L-all} is implemented using the stochastic optimiser ADAM \cite{kingma2014adam}. Each stochastic optimisation step contains a batch of $128$ images, each of which is randomly cropped into the size of $40\times 40$ pixels (and divided by $255$ if the original scale of pixel values is from 0 to 255). Following the setting of \cite{ke2021unsupervised}, we first minimise $\widehat{\mathcal{L}}$ for $2 \times 10^5$ optimisation steps with a learning rate of $0.001$ and $\alpha = 1$. This is a pretraining step for reducing the computational cost because loss $\mathcal{L}_{\rm c}$ is not evaluated at the moment. Then a further $6 \times 10^4$ optimisation steps are performed on the full loss $\mathcal{L}_{\rm all}$, with a step decay of learning rate from $0.001$ to $0.0001$ and $0.00005$. The parameter $\alpha$, in this case, is randomly selected from $[0.1,0.5]$ 

\medskip
\textit{\hbox{Loss functions.}} The input images $\{\hat{y}\}$ are generated using the given noisy images $\{y\}$ and random vectors $\{z\}$ drawn from normal distributions $\mathcal{N}(0, f^{(l)}(y))$. The generated images $\{\hat{y}\}$ are used to compute the loss function $\widehat{\mathcal{L}}$. 

To compute the loss $\mathcal{L}_{\rm c}$ (in \eqref{eq:L-all}), one needs perturbed samples $\hat{y}^{(1)}$, $\hat{y}^{(2)}$ and $\hat{y}^{(3)}$ of the noisy image $y$ (setting $P=3$). Here we let $\hat{y}^{(1)} = \hat{y}$. The sample $\hat{y}^{(2)}$ is equal to $\hat{y}$ except on a randomly selected subset $\mathcal{S}$ of pixels. For each pixel location $i$ in $\mathcal{S}$, the value $[\hat{y}^{(2)}]_i$ is set to $[\hat{y}]_{i'}$ where $i'$ is randomly selected from the $4$-neighbours of $i$ (and additionally, the associated $[z]_{i'}$ is regenerated to form $[\hat{y}^{(2)}]_i$). To avoid large perturbations in $\hat{y}^{(2)}$, the set $\mathcal{S}$ contains pixels that are separate from each other and only one pixel falls in each of the disjoint $5\times 5$ patches. Finally, the third sample is given by $\hat{y}^{(3)} = \tau_1 \hat{y}^{(1)} + (1-\tau_1) \hat{y}^{(2)}$ where $\tau_1$ is drawn from a uniform distribution in $(0,1)$. The diagonal entries of the weighting matrix $W$ are $W_{ii} = 0$ if $i \notin \mathcal{S}$ and $W_{ii} = 1/\qut{\qutb{[R(\hat{y}^{(1)})-R(\hat{y}^{(2)})]_i} + \epsilon_{\mathcal{S}} } $ if $i \in \mathcal{S}$. Here $\epsilon_{\mathcal{S}}$ is a factor for preventing division by small numbers, and it is given by $0.1 \sigma_\mathcal{S}$ where $\sigma_\mathcal{S}^2$ is the mean squared distance between $\hat{y}^{(1)}$ and $\hat{y}^{(2)}$ restricted on the subset $\mathcal{S}$. 

\medskip
\textit{\hbox{Updates of the variance of $\bmz$.}} The update of $\{\beta_1, \beta_2, \cdots, \beta_K\}$ using \eqref{eq:8615} is a crucial step of (S2). Throughout the experiments, we set $K=2$, and parameterise $f$ with polynomials of degree one. The function $f$ is initialised as a constant function based on a rough estimate of the noise variance: $\beta_1 = \hat{\sigma}^2$ and $\beta_2 = 0$. We start updating $\{\beta\}$ only after $6 \times 10^3$ optimisation steps of the full loss $\mathcal{L}_{\rm all}$, as the $R$ can be unstable at the beginning of the optimisation. 

To update $\{\beta_k\}$ requires first estimating $\delta R^\perp$. For smooth regions of the variation $\delta  R^\parallel$, we can compute $ \delta R^\perp$ by removing the smooth components of $\delta R$ (i.e., $\delta  R^\parallel$) according to the decomposition \eqref{eq:deltaR-decompose}. The details of obtaining $\delta R^\perp$ are as follows. 

First, the variation of the denoiser is computed as $\delta R = R(\hat{y}^{(2)})-R(\hat{y}^{(1)})$ where we have reused the perturbed samples $\hat{y}^{(2)}$ and $\hat{y}^{(1)}$. One of the benefits of doing so is computation effort saving because no extra forward passes of network $R$ are needed (instead, reusing  $R(\hat{y}^{(1)})$ and $R(\hat{y}^{(2)})$ that have been computed in (S1)). 

Second, $\delta R^\perp$ is computed from $\delta R$. We compute and use only a subset of pixels in $\delta R^\perp$ where $\delta R^\parallel$ is smooth (details in the next step). 
For $i$ within smooth regions of $\delta R^\parallel$, $\delta R^\perp$ is estimated by $[\delta R^\perp]_i = [{\delta R}]_i - [{\delta \tilde R}]_i$ where $[\delta \tilde{R}]_i$ denotes the average of the 4-neighbours of $\quts{\delta R}_i$. 
To find these regions, we hypothesise that in the smooth regions of $\delta R^\parallel$, $R^\parallel$ is relatively stable with respect to the changes of noise, i.e.,  the absolute values of $\delta R^\parallel$ are small relative to $\delta y$.

Third, for the region selection, we make use of the 4-neighbours of $i \in \mathcal{S}$ where $[\delta R^\perp]_j \approx [L^\perp]_{jj} [\delta y]_j= 0$ for $j$ belonging to the 4-neighbours of $i$ (recalling that by design elements in $\mathcal{S}$ are spatially separated and hence $[\delta y]_j = 0$), and consequently, $[\delta R^\parallel]_j \approx [\delta R]_j$, and also $[{\delta \tilde R}]_i \approx [{\delta \tilde R^\parallel}]_i$ ($[{\delta \tilde R^\parallel}]_i$ here means the averaged 4-neighbours of $[{\delta R^\parallel}]_i$). 
A list of pixels is selected based on $[{\delta \tilde R}]_i$ with the following steps: 
i) we divide the range of pixel values in $R(\hat{y}^{(1)})$ into $10$ intervals of equal lengths and divide $\mathcal{S}$ into $10$ subsets accordingly, based on which intervals the pixels are in. 
ii) In each subset of $\mathcal{S}$, we remove its first $20\%$ elements with the smallest values of $\qutb{\quts{\delta y}_i}$. This aims to avoid the instability of division by a small $\qutb{\quts{\delta y}_i}$ in a later step. 
iii) In each of the new subsets, we empirically select the first $3\%$ pixels with the smallest values of $\qutb{[\delta \tilde{R}]_i / [\delta y]_i}$, which correspond to small $R^\parallel$ relative to $\delta y$.
The resulting $10$ subsets of $\mathcal{S}$ are used to estimate  $\delta R^\perp$, which is then used to estimate $L^\perp$, as illustrated in subsection \ref{subsect:update-z}. 

Lastly, with the estimated $L^\perp$, the variance of $\bmz$ is updated 
by solving the least squared problem \eqref{eq:8615} restricted to the subsets of pixels where $L^\perp$ is estimated. 

\medskip
\textit{\hbox{Exponential moving average and delayed updates.}}
The parameters $\{\beta_k^{(l)}\}$ are updated within an exponential moving average (EMA) scheme \eqref{eq:8615}. The parameter $\upsilon$ controls the speed of the update. A larger $\upsilon$ takes larger steps to the estimated $\{\beta_k\}$, while a smaller $\upsilon$ might lead to slower convergent but is more robust to the errors in estimation. In our experiments we choose $\upsilon = 5 \times 10^{-4} $. 

Finally, the size of $\mathcal{S}$ must not be too small in order to have robust solutions for \eqref{eq:8615}. However, the size of $\mathcal{S}$ is restricted by the batch sizes and image sizes. A natural strategy to allow a larger sized $\mathcal{S}$ is to combine the data from multiple iterations. Specifically, we delay (S2) for a set of consecutive iterations, during which the data $\hat{y}^{(1)}$, $\hat{y}^{(2)}$, $R(\hat{y}^{(1)})$ and $R(\hat{y}^{(2)})$ are stored and merged together. The step (S2) is performed once in the last iteration, where the stored data is treated as a single batch (hence a larger size of $\mathcal{S}$). In our experiment we do one (S2) update per $5$ iterations.

\subsection{Denoising and noise variance estimation results}\label{subsect:denoising_varianceestimation} 
Following \cite{chen2016trainable, zhang2017beyond}, we use a dataset of $400$ natural images for training where each image is of size $180\times 180$. 
The denoising performance is evaluated on two different test sets, namely, BSD68 (which consists of $68$ image) \cite{martin2001database} and 12 wildly used test images Set12, following the setting of \cite{zhang2017beyond, ke2021unsupervised}. 
We consider various settings of noise (including Gaussian noise and Poisson noise at different levels). In addition to the training details given in subsection \ref{subsect:trainingdetails}, we set the weighting parameter $\gamma$ in \eqref{eq:L-all} to $1$ for Gaussian noise and to $1,4,16$ for Poisson noise with parameter $\lambda=60,30,15$, respectively. The noise variance parameters are initialised by $\beta_1 = \hat{\sigma}^2$ with $\hat{\sigma}^2 = 1/80, 1/40, 1/20$ for noise parameters $\lambda=60,30,15$, respectively. 
For each noise setting, a denoising model is trained using only the given noisy images, and their performance is reported and compared to other denoising approaches, such as BM3D \cite{dabov2007image},
Noise2Self\cite{batson2019noise2self}, 
SURE \cite{soltanayev2018training},
DPLD \cite{ke2021unsupervised}, and the supervised baseline DnCNN \cite{zhang2017beyond}. For fair comparisons, all deep learning based approaches use the same network architectures as DnCNN \cite{zhang2017beyond}.

\begin{table}[ht]
    \centering
    \renewcommand{\arraystretch}{1.2}
    \caption{
    Quality of Gaussian noise removal (measured by PSNR (dB) and SSIM) on two test sets. (the superscript $^\dagger$ means results obtained using known noise level)}\label{tab:gaussian}
\begin{tabular}{c}
(a). BSD68 \cite{martin2001database}\smallskip\\
\begin{tabular}{|c|cc||cc|}
        \toprule
        Noise Level & \multicolumn{2}{c||}{$\sigma=25$} & \multicolumn{2}{c|}{$\sigma=50$} \\
        \hline
        Measure & PSNR & SSIM & PSNR & SSIM \\
        \hline
        
        BM3D$^\dagger$\cite{dabov2007image} & $28.58$ & $0.8861$ & $25.66$ & $0.8041$ \\
        Noise2Self\cite{batson2019noise2self} & $27.48$ & $0.8588$ & $25.15$ & $0.7818$ \\
       SURE$^\dagger$\cite{soltanayev2018training} &  $28.99$ & $0.8961$ & $25.88$ & $0.8118$ %
       \\
        DPLD$^\dagger$\cite{ke2021unsupervised} & ${29.08}$ & ${0.8961}$ & ${26.13}$ & ${0.8196}$ \\
        \textbf{DVP (Ours)} & 29.06 & 0.8948 & 26.11 & 0.8168 \\
        \hline
        Supervised DnCNN \cite{zhang2017beyond} & $29.22$ & $0.9017$ & $26.24$ & $0.8265$ \\
        \bottomrule
        
        \end{tabular}\medskip\\

        (b). 12 wildly used test images \smallskip\\
        \begin{tabular}{|c|cc||cc|}
        \toprule
        Noise Level & \multicolumn{2}{c||}{$\sigma=25$} & \multicolumn{2}{c|}{$\sigma=50$} \\
        \hline
        Measure & PSNR & SSIM & PSNR & SSIM \\
        \hline
        
        BM3D$^\dagger$\cite{dabov2007image} & $29.97$ & 0.9233 & $26.71$ & 0.8663 \\
        Noise2Self\cite{batson2019noise2self} & $28.81$ & 0.9089 & $26.14$ & 0.8488 \\
       SURE$^\dagger$\cite{soltanayev2018training} &  $30.12$ & 0.9243 & $26.62$ & 0.8595   %
       \\
        DPLD$^\dagger$ \cite{ke2021unsupervised} & 30.28 & 0.9267 & 27.05 & 0.8725 \\
        \textbf{DVP (Ours)} & $30.28$ & $0.9261$ & $27.05$ & $0.8721$ \\
        \hline
        Supervised DnCNN\cite{zhang2017beyond} & $30.44$ & 0.9298 & $27.19$ & 0.8772  \\
        \bottomrule
        \end{tabular}
        \\
\end{tabular}
\end{table}

\begin{table}[ht]
    \centering
    \setlength{\tabcolsep}{2pt}
    \renewcommand{\arraystretch}{1.2}
    \caption{Estimated noise variances and their relative errors (Gaussian noise) }\label{tab:gaussian-noisevariance}
    \begin{tabular}{|c|cc|cc|}
    \toprule
 & \multicolumn{2}{c|}{$\sigma = 50$} & \multicolumn{2}{c|}{$\sigma = 25$} \\
 \hline
 & \makecell{Estimated\\ values} & \makecell{relative \\ error \\ (\%)} & \makecell{Estimated\\ values} & \makecell{relative \\ error \\ (\%)} \\
\hline
\hline
Donoho et al. \cite{donoho1994ideal}  & $3.854 \times 10^{-2}$  & 0.235  & $1.031 \times 10^{-2}$  & 7.291  \\
Immerkaer \cite{immerkaer1996fast} & $4.227 \times 10^{-2}$  & 9.935  & $1.128 \times 10^{-2}$  & 17.318  \\
Foi et al. \cite{foi2008practical} & $4.029 \times 10^{-2}$  & 4.788  & $1.058 \times 10^{-2}$  & 10.056  \\
Rakhshanfar et al. \cite{rakhshanfar2016estimation} & $3.886 \times 10^{-2}$  & 1.063  & $0.983 \times 10^{-2}$  & 2.271  \\
Pimpalkhute et al. \cite{pimpalkhute2021digital} & $3.866 \times 10^{-2}$  & 0.562  & $0.974 \times 10^{-2}$  & 1.386  \\
DVP (ours) & $\bm{3.841 \times 10^{-2}}$  & ${-0.093}$  & $\bm{0.970 \times 10^{-2}}$  & ${0.931}$  \\
\hline
Ground truth & $3.845 \times 10^{-2}$  & -  & $0.961 \times 10^{-2}$  & -   \\
\bottomrule
    \end{tabular}
\end{table}

\subsubsection{Gaussian white noise}
We consider two different noise levels with standard deviations of $\sigma = 30$ and $60$ respectively (relative to image pixel values ranging from $0$ to $255$). For each noise level, the results, measured in both the Peak signal-to-noise ratio (PSNR) and the structural similarity (SSIM) index, are presented in Table \ref{tab:gaussian}. In comparison with BM3D \cite{dabov2007image}, one of the best performing model-based Gaussian denoisers, our method (DVP) has a higher PSNR (by around $0.5$ dB) for both noise levels and both test sets. Also, DVP performs on par with the unsupervised method DPLD \cite{ke2021unsupervised} and outperforms SURE \cite{soltanayev2018training} (by around $1$ to $1.5$ dB in all cases). It is worth pointing out that the results of BM3D, SURE, and DPLD are obtained using the ground truth noise levels (in contrast, our approach does not require such information). Remarkably, DVP reaches a denoising quality similar to that of the supervised baseline DnCNN \cite{zhang2017beyond} (trained using noisy images and ground truth pairs), with a gap of around $0.1$ or $0.2$ dB. 

\begin{figure*}[ht]
    \centering
    \setlength{\tabcolsep}{5pt}
    \begin{tabular}{c|ccc}
    \includegraphics[width=0.22\textwidth]{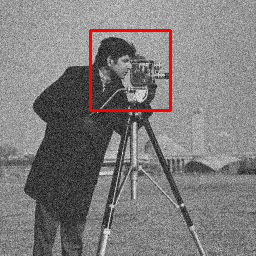} & 
    \includegraphics[width=0.22\textwidth]{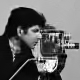} &
    \includegraphics[width=0.22\textwidth]{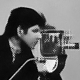} &
    \includegraphics[width=0.22\textwidth]{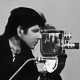}
    \\
    {\small Noisy image} & {\small BM3D$^\dagger$ \cite{dabov2007image}} & {\small Noise2Self \cite{batson2019noise2self}} & {\small SURE$^\dagger$ \cite{soltanayev2018training}} 
    \\
    \includegraphics[width=0.22\textwidth]{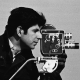} &
    \includegraphics[width=0.22\textwidth]{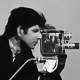} &
    \includegraphics[width=0.22\textwidth]{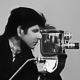} &
    \includegraphics[width=0.22\textwidth]{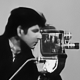} 
    \\
    {\small Ground truth} & {\small Supervised DnCNN \cite{zhang2017beyond}} & {\small DPLD$^\dagger$ \cite{ke2021unsupervised}} & {\small DVP (Ours)}\\
    \end{tabular}
    \caption{Gaussian denoising results ($\sigma=25$). The images on the last three columns reflect the denoising results by different methods for the region of the noisy image highlighted in red square. The symbol $^\dagger$ means that the ground truth noise level is used.}
    \label{fig:naturalimage-results}
\end{figure*}

Our approaches generated noise variance estimates during the learning process (from the noisy training images). The results are summarised in Table \ref{tab:gaussian-noisevariance}, where the relative error means $(\sigma^2_{\rm estimated}\!-  \!\sigma^2)/\sigma^2$ for an estimated variance $\sigma^2_{\rm estimated}$. Our approach is not specific to Gaussian noise, and in this case, we simply set $\sigma^2_{\rm estimated} = \beta_1$ (where $\beta_1$ is the coefficient of the constant term of the computed $f$ at the last iteration). The performance is compared to that of the other noise variance estimation methods, for each of which the results are obtained by averaging its estimated noise variances over the $400$ training images. The table shows relative errors of less than $1\%$ of our approach in both noise levels, and they are significantly smaller than that of the other methods. The obtained accurate noise variances are in good accordance with the similar denoising quality between DVP and DPLD \cite{ke2021unsupervised} given in Table \ref{tab:gaussian}, where the latter uses ground truth noise variance. 

The qualitative results are presented in Figure \ref{fig:naturalimage-results}, in which the denoised images by different methods for a standard test image \textit{Cameraman} under Gaussian noise with $\sigma = 25$ are compared. The last three columns are denoised results specific to the region of the noisy image that is highlighted by the red square (on the top-left of the figure). Our method is able to capture the smooth parts as well as the sharp edges of the images (as shown on the bottom-right of the figure), and it achieves a similar visual quality to DPLD \cite{ke2021unsupervised}. By comparison, the BM3D method also recovers well the fine details but small artefacts are observed around the objects in the denoised images. The result of the Noise2Self method, in contrast, looks less smooth compared to the others.

\begin{table}[ht]
    \centering
    \renewcommand{\arraystretch}{1.2}
    \setlength{\tabcolsep}{4pt}
    \caption{
    Quality of Poisson noise removal (measured by PSNR (dB) and SSIM) on two test sets. (the superscript $^\dagger$ means results obtained using known noise level)}\label{tab:poisson}
\begin{tabular}{c}
(a). BSD68 \cite{martin2001database}\smallskip\\
\begin{tabular}{|c|cc||cc||cc|}
        \toprule
        Noise parameters  & \multicolumn{2}{c||}{$\lambda=60$} & \multicolumn{2}{c||}{$\lambda=30$} & \multicolumn{2}{c|}{$\lambda=15$} \\
        \hline
        Measure & PSNR & SSIM & PSNR & SSIM & PSNR & SSIM \\
        \hline
        Noise2Self\cite{batson2019noise2self} & $27.78$ & $0.8660$ & $26.56$ & $0.8285$ & $25.42$ & $0.7902$  \\
        DPLD$^\dagger$\cite{ke2021unsupervised} & ${29.28}$ & ${0.9018}$ & ${27.65}$ & ${0.8625}$ & ${26.16}$ & ${0.8210}$  \\
        \textbf{DVP (Ours)} & 29.26 & 0.9009 & 27.67 & 0.8647 & 26.18 & 0.8210 \\
        \hline
        \makecell{Supervised \\ DnCNN\cite{zhang2017beyond}}  & $29.43$ & $0.9081$ & $27.86$ & $0.8743$ & $26.39$ & $0.8348$ \\
        \bottomrule
        
        \end{tabular}\medskip\\

        (b). 12 wildly used test images \smallskip\\
        \begin{tabular}{|c|cc||cc||cc|}
        \toprule
        Noise parameters & \multicolumn{2}{c||}{$\lambda=60$} & \multicolumn{2}{c||}{$\lambda=30$} & \multicolumn{2}{c|}{$\lambda=15$} \\
        \hline
        Measure & PSNR & SSIM & PSNR & SSIM & PSNR & SSIM\\
        \hline
        Noise2Self\cite{batson2019noise2self} & $29.15$ & 0.9100 & $27.82$ & 0.8847 & $26.26$ & 0.8439 \\
        DPLD$^\dagger$ \cite{ke2021unsupervised} & ${30.34}$  & 0.9268 & ${28.68}$  & 0.9019 & ${26.99}$ & 0.8698 \\
        \textbf{DVP (Ours)} & 30.33 & 0.9269 & 28.72 & 0.9042 & 27.04 & 0.8722 \\
        \hline
        \makecell{Supervised \\ DnCNN\cite{zhang2017beyond}} & $30.49$ & 0.9306 & $28.87$ & 0.9086 &  $27.28$ & 0.8801  \\
        \bottomrule
        \end{tabular}
        \\
\end{tabular}

\end{table}

\subsubsection{Poisson noise}
The denoising results are compared in Table \ref{tab:poisson}. Similar to the Gaussian noise cases, the proposed method DVP has PSNR and SSIM scores comparable to that of DPLD \cite{ke2021unsupervised} (which uses ground truth noise parameters) for the different values of parameter $\lambda$ and for both test sets (the BSD68 \cite{martin2001database} and the 12 wildly used test images). Moreover, the gap between DVP and the supervised baseline is around $0.1$ to $0.2$ dB in PSNR in all reported cases, which is much smaller than the gap for Noise2Self (from around $1$ to $1.5$ dB). 

Noise distributions here are more complex compared to the Gaussian cases as they are signal dependent. Our method works also for signal-dependent noise, as the noise variance estimation results in Table \ref{tab:poisson-noisevariance} show. In the table, MAE (mean absolute errors) and RMAE (relative mean absolute errors) are defined respectively by 
\[
\begin{split}
    & \sum_{l=0}^{255} \qutb{ [v_{\rm est}]_l - [v_{\rm GT}]_l}  \ {\rm and} \ \left. \sum_{l=0}^{255} \qutb{ [v_{\rm est}]_l - [v_{\rm GT}]_l} \middle/ \sum_{l=0}^{255} [v_{\rm GT}]_l \right.,
\end{split}
\]
where $[v_{\rm est}]_l$ is the estimated variance at intensity level $l$ (ranging from $0$ to $255$), and $[v_{\rm GT}]_l$ is the associated ground truth variance.
As a comparison, we report in the table the results of two other noise variance estimation methods on the same dataset ($400$ noisy images). 
Our method DVP has RMAE around $5$ times smaller than that of the other two methods for the three different values of $\lambda$. 

\begin{table}[ht]
    \centering
    \setlength{\tabcolsep}{3pt}
    \caption{Noise variance estimation results for Poisson noise. MAE and RMAE refer to mean absolute errors and relative mean absolute errors, respectively}
    \label{tab:poisson-noisevariance}
    \renewcommand{\arraystretch}{1.3}
    \begin{tabular}{|c|cc|cc|cc|}
    \toprule
    & \multicolumn{2}{|c|}{$\lambda = 60$} &  \multicolumn{2}{c|}{$\lambda = 30$} &  \multicolumn{2}{c|}{$\lambda = 15$} \\
    \hline
     & \makecell{ MAE \\ $\times 10^3$ } & \makecell{RMAE\\ ($\%$)} & \makecell{ MAE \\ $\times 10^3$ } & \makecell{RMAE\\ ($\%$)} & \makecell{ MAE \\ $\times 10^3$ } & \makecell{RMAE\\ ($\%$)} \\
    \hline
    \hline
Foi et al. \cite{foi2008practical} & $1.695$  & $4.238 $ & $1.317$  & $6.585 $ & $1.003$  & $10.026 $  \\
Rakhshanfar et al. \cite{rakhshanfar2016estimation} & $1.857$  & $4.643 $ & $0.957$  & $4.787 $ & $0.647$  & $6.468 $  \\
DVP (Ours) & $\bm{0.260}$  & $\bm{0.650}$ & $\bm{0.149}$  & $\bm{0.745}$ & $\bm{0.147}$  & $\bm{1.469}$  \\
\bottomrule
    \end{tabular}
\end{table}

\subsection{Denoising real microscopy images.}\label{subsect:microscopyimages}
By design, our algorithm is suitable for learning denoising from noisy images and without prior knowledge about the noise variance. We apply the algorithm to real microscopy images, where the ground truth images and noise variance are unknown. Following the setting of \cite{ke2021unsupervised}, we consider two fluorescence microscopy datasets, namely, N2DH-GOWT1 (images of GFP transfected GOWT1 mouse stem cells) and C2DL-MSC (images of rat mesenchymal stem cells), each of which is a video composed of a number of noisy frames.  The first dataset N2DH-GOWT1 contains $92$ frames of the size $1024 \times 1024$, while the second one has $48$ frames of size $992 \times 832$. The denoisers are trained on these noisy frames only. 
In addition to the training details given in subsection \ref{subsect:trainingdetails}, the weighting parameter $\gamma$ in \eqref{eq:L-all} is set to $1$, and the noise variance parameters are initialised by $\beta_1 = 0.01$ and $\beta_2=0$. 

\begin{figure*}[ht]
    \centering
    \setlength{\tabcolsep}{2pt}
    \makebox[\textwidth][c]{
    \begin{tabular}{cccccc}
    \raisebox{-.1\height}{\rotatebox{90}{\textbf{(a). N2DH-GOWT1}}} &
    \includegraphics[width=0.2\textwidth]{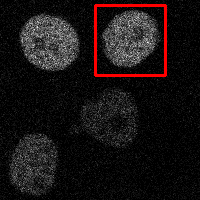} &
    \includegraphics[width=0.2\textwidth]{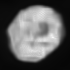} &
    \includegraphics[width=0.2\textwidth]{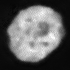} & 
    \includegraphics[width=0.2\textwidth]{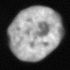} &
    \includegraphics[width=0.2\textwidth]{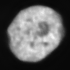}  \\    
     & {\small Noisy image} & {\small BM3D \cite{dabov2007image}} & {\small Noise2Self \cite{batson2019noise2self}} & {\small DPLD \cite{ke2021unsupervised}} & {\small DVP (Ours)}  \\
    \\
    \raisebox{-.0\height}{\rotatebox{90}{\textbf{(b). C2DL-MSC}}} &
    \includegraphics[width=0.2\textwidth]{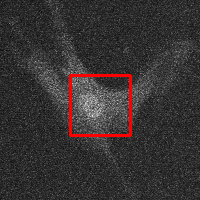} &
    \includegraphics[width=0.2\textwidth]{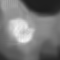} &
    \includegraphics[width=0.2\textwidth]{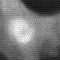} & 
    \includegraphics[width=0.2\textwidth]{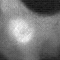} &
    \includegraphics[width=0.2\textwidth]{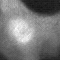}  \\    
    & {\small Noisy image} & {\small BM3D \cite{dabov2007image}} & {\small Noise2Self \cite{batson2019noise2self}} & {\small DPLD \cite{ke2021unsupervised}} & {\small DVP (Ours)}  \\
    \end{tabular}}
    \caption{Denoising results for microscopy images N2DH-GOWT1 (top row) and C2DL-MSC (bottom row). The images displayed in the first column are noisy images (i.e., inputs to denoisers). The other columns correspond to denoised results obtained by different methods on a sub-region of the noisy images highlighted by the red square in the first column.}
    \label{fig:microscopy-results}
\end{figure*}

Examples of denoised results by the proposed approach are given in Figure \ref{fig:microscopy-results}. The results are obtained based on the same noisy frames of the videos that were used for training (as our method can be trained directly on the targeted images that need to be denoised). Samples of the noisy frames are displayed in the first column of the images, where the red squares highlight sub-regions for which the denoised results by different methods are demonstrated (see the last 4 columns). As shown in the last column of Figure \ref{fig:microscopy-results}, our approach recovers well the smooth regions of the images, despite the fact that such regions are not seen by the training process. The denoised results are visually close to that of DPLD \cite{ke2021unsupervised}, but we point out again that the latter requires a separate noise variance estimation step. Furthermore, in the figure, we also present the results of BM3D \cite{dabov2007image} and Noise2Self \cite{batson2019noise2self} as baselines. Our method is able to reveal more visual structure details of the underlying objects compared to the results of BM3D (second column) and Noise2Self (third column).

\section{Conclusion}
In this work, we propose an unsupervised deep learning method for jointly learning denoisers and estimating noise variance from a set of noisy images. Our method is based on deep variation priors, which state that the variation of denoisers with respect to the changes of noise follows some smoothness assumptions. We present deep variation priors as a criterion for high-quality denoisers in a data-driven setting, and such priors are combined with deep neural networks to explain a set of noisy training images, without having any ground truth images or noise variance beforehand. By doing so, our method achieves high-quality denoisers as well as accurate noise variance estimation. We demonstrate the power of the proposed method in denoising real microscopy images, where paired noisy and clean training images do not exist. 

The proposed learning model is end-to-end, in contrast to other unsupervised denoising solutions that require a post-processing step or a separate process of noise estimation. In particular, our method leverages the error information of denoising for better noise variance estimation, which in return improves the quality of denoisers.

\section*{Acknowledgments}
The experiments in this work were carried out with support from the computational facilities of the Advanced Computing Research Centre, University of Bristol. %

{\appendices
\section{Approximating $M$}\label{sect:appendixA}

We aim to show that,  
\[
R^\perp(\bmyh) \approx g^\perp(\bmx) + L^\perp(\bmn + \alpha \bmz),
\]
where 
\begin{equation}
    L^\perp = (I-M) (I + \alpha^2 M)^{-1}. \tag{\ref{eq:LM}}
\end{equation}
The above formulation has been used in subsection \ref{subsect:update-z}. 

\medskip
For simplicity, we assume that the condition of Theorem 1 holds. The desired formulation can be derived in the following steps.  %

(i). Denote by $\mathcal{C}$ the set of denoisers satisfying \eqref{eq:pld} with $\bmeh = 0$. For any $R^0 \in \mathcal{C}$, we rewrite the loss function \eqref{eq:th1} as 
\begin{equation}\label{eq:5714}
\begin{split}
    \widehat{\mathcal{L}}(R^0) & = 
\bbE \qut{\qutn{ R^\parallel\qut{\bmyh} - \bmx +  \qut{R^{0\perp}\qut{\bmyh} - (I-M)\bmn }  }^2} + c \\
& = \bbE\qut{\qutn{\bmp}^2} + 2 \bbE\qut{\qutan{\bmp,\bmq}} + \bbE\qut{\qutn{\bmq}^2} + c,
\end{split}
\end{equation}
where $R^{0\perp}:=R^0-R^\parallel$, $\bmp:=R^\parallel\qut{\bmyh} - \bmx$ and $\bmq:=R^{0\perp}\qut{\bmyh} - (I-M)\bmn$. 

(ii).  Next we show that the cross term $\bbE\qut{\qutan{\bmp,\bmq}}$ is independent of $R^{0\perp}$. Since $R^\parallel, R^0 \in \mathcal{C}$, one has $R^{0\perp} \in \mathcal{C}$. Moreover, by definition, $R^\parallel$ is the minimiser of $\bbE\qut{\qutn{ R^1\qut{\bmyh} - \bmx}^2}$ over all $R^1 \in \mathcal{C}$. The optimality of $R^\parallel$ implies that 
\[
\bbE\qut{\qutan{ R^\parallel\qut{\bmyh} - \bmx,  R^{0\perp}\qut{\bmyh} }} = 0,
\]
for any $R^{0\perp} \in \mathcal{C}$. Therefore, $\bbE\qut{\qutan{\bmp,\bmq}} = \bbE\qut{\qutan{\bmp,  (I-M)\bmn}}$ and it does not depend on $R^{0\perp}$. 

(iii). By Equations \eqref{eq:5714} and the observation in (ii), 
the minimisation of $\widehat{\mathcal{L}}(R^0)$ is equivalent to 
\[
\min_{R^0 \in \mathcal{C}} \widehat{\mathcal{L}}(R^0) = \bbE\qut{\qutn{\bmp}^2} + 2 \bbE\qut{\qutan{\bmp,\bmq}} + \min_{R^{0\perp} \in \mathcal{C}} \bbE\qut{\qutn{\bmq}^2} + c.
\]
Therefore, writing the minimiser of $\widehat{\mathcal{L}}$ as $R^* = R^\parallel + R^\perp$ (Cf. Equation \eqref{eq:7512}), the term $R^\perp$ should satisfy
\begin{equation}\label{4152}
    \bbE\qut{\qutn{R^{\perp}\qut{\bmyh} - (I-M)\bmn}^2} = 
\min_{R^{0\perp} \in \mathcal{C}}  \bbE\qut{\qutn{\bmq}^2}.
\end{equation}

(iv). Since $R^\perp \in \mathcal{C}$, it can be written in the following form
\[
R^\perp(\bmyh) = g^\perp(\bmx) + L^\dagger(\bmn + \alpha \bmz).
\]
Here $L^\dagger$ depends on $\bmx$. Plugging this into \eqref{4152}, $L^\dagger$ approximately minimises   
\[
\bbE\qut{\qutn{g^\perp(\bmx) + L^{\dagger}(\bmn + \alpha \bmz) - (I-M)\bmn}^2},
\]
Therefore, 
\[
\begin{split}
    L^\dagger \approx  L^\perp & :=  (I-M) \cov{\bmn \mid \bmx} \qut{\cov{\bmn \mid \bmx} + \alpha^2 \cov{\bmz \mid \bmx}}^{-1} \\
    & = (I-M) (I + \alpha^2 M)^{-1}. 
\end{split}
\]
which gives the desired statement.

}

\bibliographystyle{plain}
\bibliography{references}

\end{document}